\documentclass[10pt,twocolumn,letterpaper]{article}

\usepackage{cvpr}              
\usepackage{flushend}
\usepackage{balance}

\usepackage[dvipsnames]{xcolor}

\definecolor{cvprblue}{rgb}{0.21,0.49,0.74}
\usepackage[pagebackref,breaklinks,colorlinks,citecolor=cvprblue]{hyperref}
\newcommand{\RNum}[1]{\uppercase\expandafter{\romannumeral #1\relax}}

\def\paperID{*****} 
\def\confName{CVPR}
\def\confYear{2024}

\title{How Suboptimal is Training rPPG Models with Videos and Targets from Different Body Sites?}

\author{Björn Braun\\
Department of Computer Science\\
ETH Zürich, Switzerland\\
{\tt\small bjoern.braun@inf.ethz.ch}
\and
Daniel McDuff\\
Paul G. Allen School of Computer Science\\
University of Washington\\
{\tt\small dmcduff@uw.edu}
\and
Christian Holz\\
Department of Computer Science\\
ETH Zürich, Switzerland\\
{\tt\small christian.holz@inf.ethz.ch}
}

\begin{document}
\maketitle
\begin{abstract}
Remote camera measurement of the blood volume pulse via photoplethysmography (rPPG) is a compelling technology for scalable, low-cost, and accessible assessment of cardiovascular information. Neural networks currently provide the state-of-the-art for this task and supervised training or fine-tuning is an important step in creating these models. However, most current models are trained on facial videos using contact PPG measurements from the fingertip as targets/labels. One of the reasons for this is that few public datasets to date have incorporated contact PPG measurements from the face.  Yet there is copious evidence that the PPG signals at different sites on the body have very different morphological features. 
Is training a facial video rPPG model using contact measurements from another site on the body suboptimal? 
Using a recently released unique dataset with synchronized contact PPG and video measurements from both the hand and face, we can provide precise and quantitative answers to this question. We obtain up to 40\% lower mean squared errors between the waveforms of the predicted and the ground truth PPG signals using state-of-the-art neural models when using PPG signals from the forehead compared to using PPG signals from the fingertip. We also show qualitatively that the neural models learn to predict the morphology of the ground truth PPG signal better when trained on the forehead PPG signals. However, while models trained from the forehead PPG produce a more faithful waveform, models trained from a finger PPG do still learn the dominant frequency (i.e., the heart rate) well.
\end{abstract}    
\section{Introduction}
\label{sec:intro}

Camera measurement of photoplethysmography (PPG), or remote PPG (rPPG), is compelling as it opens the door for scalable, low-cost, and comfortable passive measurement of cardiovascular information~\cite{mcduff2023camera} such as the blood volume pulse (BVP) waveform~\cite{blazek2000near,takano2007heart,verkruysse2008remote}.
The PPG signal contains various vital pieces of information about the physiological state of a subject, most notably pulse rate~\cite{poh2010non}, but also respiration rate~\cite{poh2010advancements} and blood pressure correlates~\cite{schrumpf2021assessment,jeong2016introducing,bousefsaf2022estimation,curran2023camera}.

\begin{figure}
  \centering
    \includegraphics[width=0.48\textwidth]{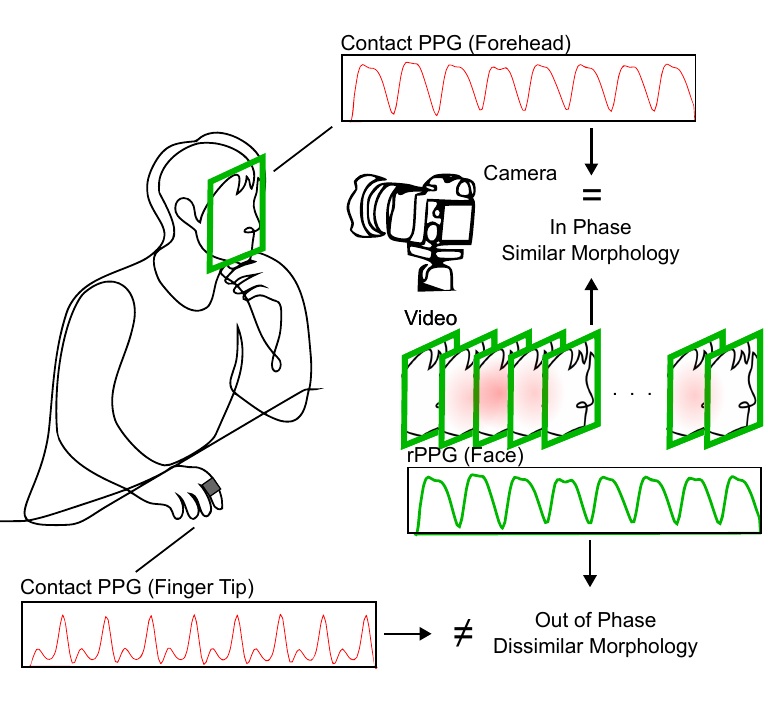}
  \caption{\textbf{What is the Optimal Target for rPPG?} Most rPPG models are trained with PPG targets from the finger tip; however, waveform morphology differs in phase and morphology at different sites across the body. We show that using finger tip PPG targets is less optimal than using PPG measured from the face.}
  \label{fig:overview}
\end{figure}

Neural models can be trained to create a mapping between pixels in a video and the BVP~\cite{chen2018deepphys,vspetlik2018visual}. 
While unsupervised learning is a popular choice~\cite{gideon2021way,yang2022simper}, these models still usually need supervised fine-tuning to achieve state-of-the-art performance. 
The best models are currently trained using deep neural networks with relatively large numbers of parameters that can learn high-dimensional, complex mappings between pixel values over time and the pulse waveform.
Almost exclusively, these supervised rPPG models are trained using target measurements from a finger tip PPG sensor and videos of peoples' faces. 
However, it has not yet been evaluated how the location of the ground truth PPG sensor influences the performance of these supervised rPPG models.
PPG signals vary significantly in morphology and phase (due to differences in pulse arrival time (PAT)) across the body~\cite{nilsson2007combined,hernando2017finger, hernando2019finger,niu2023full} (see Fig.~\ref{fig:overview}).
Therefore, the performance of the trained neural methods for rPPG measurement likely exhibits location-dependent variations, given the substantial differences in PPG signal characteristics across body sites.

More generally in machine learning tasks reducing the domain gap between the target signal and the labels is advantageous. 
By requiring the model to extract a PPG signal from face pixels and then map the facial PPG morphology to that from the finger we are making the learning task more difficult. Yet, very little prior work has considered the impact of this choice. 

In this paper, we present a comprehensive study evaluating the performance of three commonly employed rPPG models --- DeepPhys~\cite{chen2018deepphys}, TS-CAN~\cite{liu2020multi}, and PhysNet~\cite{yu2019remote} --- when trained using PPG signals from two different body locations: the forehead and the finger. 
Our findings reveal that the waveform prediction performance of these networks can be improved substantially when using PPG signals from the forehead compared to those from the finger when the models are trained with videos from the face.
The mean squared error (MSE) between the predicted and the reference contact PPG signals is improved by up to 40\% when training and testing with the forehead PPG signal compared to using the finger PPG signal.
In addition, we qualitatively compare the predicted waveforms and see that we achieve a better temporal alignment and preservation of morphological characteristics training and testing on the forehead PPG signal.
We hypothesize that this improvement is attributed to the smaller domain gap when using videos from the face as input and labels from the face, compared to using labels from the finger.
We believe that these insights into location-dependent variations will foster advancements in rPPG research, enabling the development of more accurate and robust models for physiological signal estimation and will help inform the design of future data collections.

\section{Related work}
\label{sec:related_work}

\subsection{Camera-based Physiological Measurements}
\label{sec:camera_measurements}
The field of remote photoplethysmography (rPPG) has witnessed significant advancements in recent years. 
By leveraging video cameras, such as regular webcams or smartphone cameras, rPPG enables non-contact estimation of physiological signals, such as heart rate and blood volume changes~\cite{mcduff2023camera}. 
By analyzing light absorption variations caused by blood flow changes, rPPG enables convenient vital sign assessment from optical measurements of the skin without physical sensor contact~\cite{blazek2000near, takano2007heart, verkruysse2008remote, poh2010non}.
In comparison to traditional wearable sensors such as smartwatches, this method offers increased comfortability for the user and the potential for easier scalability.

Originally, unsupervised signal processing approaches such as blind source separation~\cite{poh2010non} or mathematical models of the skin properties~\cite{wang2017algorithmic} were used to estimate the rPPG signal. Improvements have been achieved using deep learning approaches that can model more complex spatial-temporal. Examples of networks specifically designed for rPPG include DeepPhys~\cite{chen2018deepphys}, TS-CAN~\cite{liu2020multi} and PhysNet~\cite{yu2019remote}. These networks are typically trained on videos of peoples' faces as the skin is least likely to be obscured by clothing at that location. 
Furthermore, it is not only possible to estimate the BVP remotely, but also other physiological vitals such as blood pressure~\cite{jeong2016introducing, bousefsaf2022estimation}, respiratory rate (RR)~\cite{tarassenko2014non, liu2020multi}, electrodermal activity correlates~\cite{shastri2012perinasal, bhamborae2020towards}, or sympathetic arousal~\cite{braun2023video,mcduff2020non}.
Blood pressure estimation via pulse wave analysis (PWA)
use features about the shape (or morphology) of the PPG signal and therefore it is important that waveform does not only capture the pulse frequency but also resembles the ground truth PPG in order, more subtle ways.

\subsection{PPG Signal Source}
\label{sec:ppg_signal_source}
While state-of-the-art neural models have achieved the best performance, most of the neural models were trained on similar datasets such as PURE~\cite{stricker2014non}, UBFC-rPPG~\cite{bobbia2019unsupervised}, UBFC-Phys~\cite{meziatisabour2021ubfc}, AFRL~\cite{estepp2014recovering}, MMSE-HR~\cite{zhang2016multimodal}, or VIPL-HR~\cite{niu2019vipl}, which all collect the ground truth PPG signals from the finger or the wrist using reflective, contact PPG sensors.
To the best of our knowledge, none of these datasets have ground truth PPG signals obtained from sites on the face, such as the forehead.

Previous works have analyzed how the PPG signals collected from different sites of the body differ.
\citeauthor{nilsson2007combined}~\cite{nilsson2007combined} compared PPG signals from five different locations (forearm, finger, forehead, wrist, and shoulder) to evaluate the spectral power of the pulse and respiration components.
They found that the finger has significantly lower respiration spectral power than the forehead but higher pulse spectral power than the forehead.
\citeauthor{kim2021assessment} found that the region on the face, from which rPPG is predicted, influences the rPPG accuracy.
As the skin thickness is not the same at different locations on the body, the absorption properties of the skin change, influencing the specular reflections of the light.

It has, however, not yet been evaluated how the location of the ground truth PPG signal influences the performance of supervised rPPG neural models.
As the morphological features and pulse spectral power of the PPG signals are inherently different at different body sites, we hypothesize that these characteristics also influence the performance of the trained neural models.

\section{Methods}
\label{sec:methods}

\subsection{Data}
\label{sec:data}
\begin{figure}[h]
  \centering
    \includegraphics[width=0.48\textwidth]{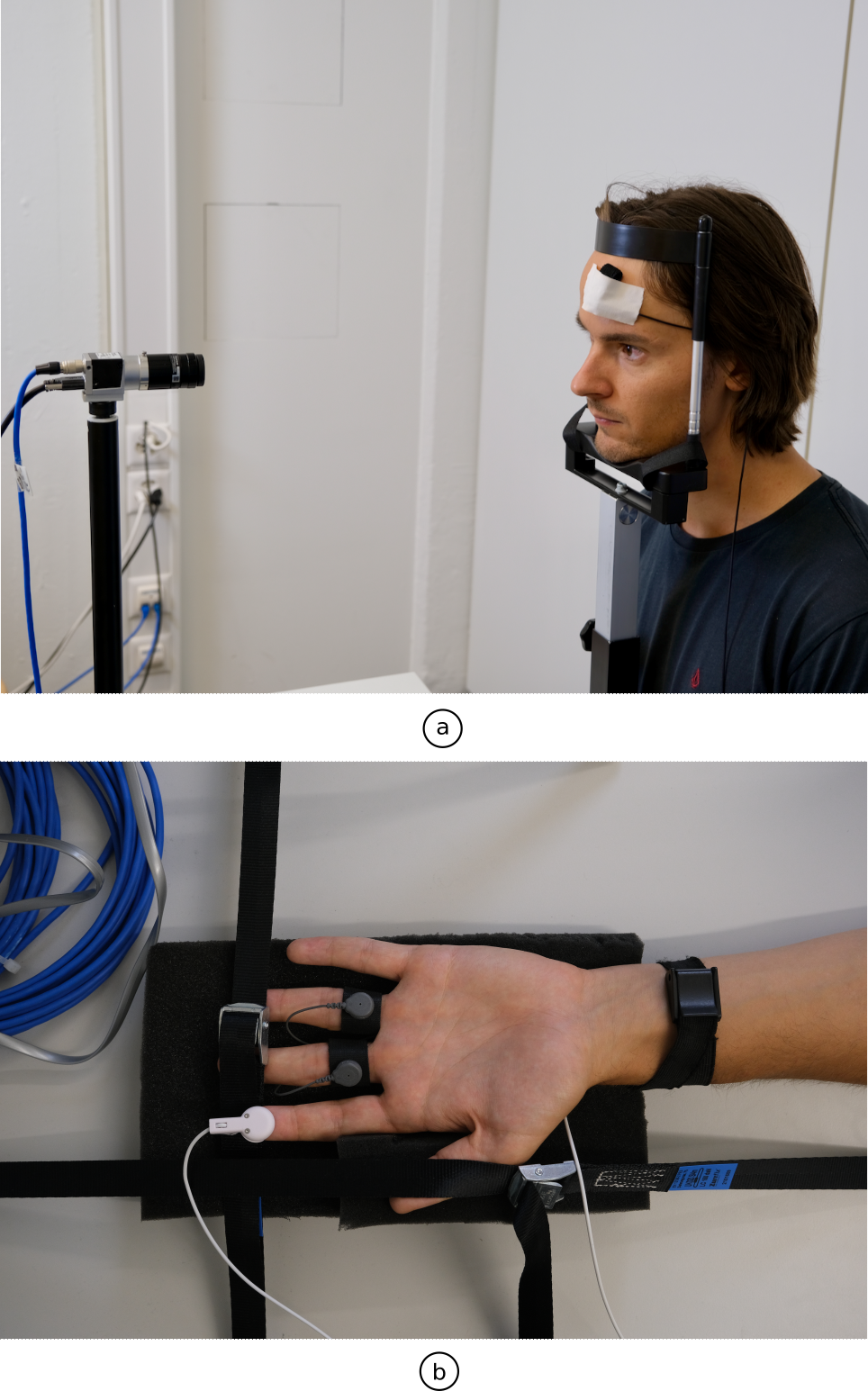}
  \caption{\textbf{Experimental Apparatus.} We used a customarily designed data collection apparatus that simultaneously collects contact reflectance PPG measurements from the face (forehead) (a) and finger (b) synchronized with video recordings of the face.}
  \label{fig:study_setup}
\end{figure}

We use a dataset of $N=18$ participants (4 female, 14 male, ages 19--36, $\mu=27.1$ and $\sigma=3.7$) for our experiments~\cite{braun2023video}. 
The unique aspect of this dataset is the synchronized collection of contact PPG and video measurement from \emph{both} the hand and face. 
We show the study setup in~\autoref{fig:study_setup}.
These data are also released publicly for future research.

Based on the Fitzpatrick scale~\citep{fitzpatrick1988validity}, 3 participants had skin type \RNum{2}, 9 skin type \RNum{3}, 3 skin type \RNum{5}, and 2 skin type \RNum{6}.
The dataset captures 9.5~minutes of video recording (with disabled white balancing, auto-focus, and auto-exposure) of participants' faces and hands, with synchronized EDA and PPG recordings from their fingers.
We recorded the videos using two Basler acA1300-200uc cameras with a recording frame rate of 100\,Hz and the physiological signals from a synchronized Shimmer3 GSR+ device and a BIOPAC MP160 that triggered the cameras. 
The Shimmer 3 GSR+ device recorded with a sampling frequency of 100\,Hz and the BIOPAC MP160 with a sampling frequency of 2000\,Hz.
The participant placed their head on a chin rest to minimize motion artifacts, and the lighting and temperature in the room was kept constant throughout the study to minimize any other confounders.
The protocol alternated between periods of resting (2\,minutes) and periods of physical stress during which the participants pinched their skin (self-pinching) (30\,seconds) to stimulate an EDA response, starting with a period of rest.

\subsection{Task Description}
The goal of our evaluation is to determine whether supervised neural models, which estimate a person's rPPG signal, achieve better performance when trained with contact PPG signals from the forehead than with PPG signals from the finger as ground truth labels.
Input into our models is the differences between the standardized camera images from our used dataset (downsampled to 25\,Hz, 72 x 72).
We train our method in a supervised setting using the differences of the contact ground truth PPG signals as labels.

\subsection{Implementation}
\label{sec:implementation}
To highlight how the results generalize, we performed experiments with three popular neural models (DeepPhys~\cite{chen2018deepphys}, TS-CAN~\cite{liu2020multi}, PhysNet~\cite{yu2019remote}).  
For the experiments, we used the open-source rPPG Toolbox~\cite{liu2024rppg}. 
By using a public dataset and the open toolbox, we ensure that our results are easily reproducible.

\subsubsection{Processing Pipeline}
We process the camera images from the dataset in four steps.
First, we crop the camera images to only include the forehead using a fixed bounding box as described by~\cite{braun2023video}.
As the heads' of the study participants of our used dataset are fixed on a chin rest, no dynamic detection of the forehead is necessary.
In the second step, we resize the images to a resolution of 72x72.
Then, we downsample the frame rate of the videos to a sampling frequency of 25\,Hz, which is high enough to capture the typical frequency band of the heart rate (0.7--2.5\,Hz~\citep{spodick1992operational, liu2020multi}).
Finally, we take the consecutive difference between the frames and standardize them by dividing them through the standard deviation of the pixel intensity values as it was proposed by \citeauthor{chen2018deepphys}~\cite{chen2018deepphys} and originally used for the DeepPhys~\cite{chen2018deepphys} and TS-CAN~\cite{liu2020multi} models. 
Because we used a frame rate of 25\,Hz and the standardized frame difference as an input into our neural model, we also use a sampling rate of 25\,Hz and the differences of the PPG signals as labels for our models.
As the Shimmer GSR3+ device inherently filters out very low-frequent components of the rPPG signal, we filter the PPG signal from the BIOPAC device with a 2\textsuperscript{nd} order Butterworth high pass filter with a cutoff frequency of 0.7\,Hz to allow for a fair comparison of both signals.

\subsubsection{Training}
Our model training employs the LOSO cross-validation strategy across all 18 participants to ensure robust evaluation across subjects.
In each iteration, we reserve one participant as the test set, another as the validation set, and utilize the remaining participants for training.
We train for 30 epochs until the convergence of the validation loss and set the batch size to 4 with an input window size of 256 frames.
During training, we shuffle the order of the input windows randomly to ensure that the model cannot learn any possible underlying patterns.
As the loss function, we use the mean squared error (MSE) loss for all models. 
We optimize the network using the Adam optimizer with hyperparameters $\beta_1=0.9$, $\beta_2=0.999$~\citep{kingma2014adam} and a learning rate of 0.001, dynamically adjusted using the OneCycle learning rate policy~\citep{smith2019super}.
To train the model, we use an NVIDIA GeForce RTX 4090 GPU, with a total runtime of four to nine hours (depending on the model) for all 18 participants using the LOSO cross-validation approach.

\subsubsection{Evaluation}
First, we take the cumulative sum of the predicted signals and the labels as we use the consecutive differences of the inputs/ labels during training.
We evaluate our predicted rPPG signals using the mean squared error (MSE) between the waveforms of the predicted and the ground truth signal.
This allows us to evaluate how accurately the network learned to predict the actual waveform of the ground truth PPG signal.
Before calculating the MSE, we normalize the predicted and ground truth signals between zero and one to allow for a fair comparison.
In addition, we also qualitatively compare the the predicted and ground truth PPG signals from the forehead and the finger to analyze the model performance.
Furthermore, to ensure that our trained models achieve state-of-the-art performance, we also calculate the mean absolute error (MAE) of the heart rate as this is the most commonly used metric in literature to compare the performance of trained rPPG models.
We report it for the scenario of training and testing using the face PPG as a reference.
We obtain the MAE by evaluating the heart rates with the same 30-second sliding window approach with no overlap as explained in the rPPG toolbox~\cite{liu2024rppg}.
First, we filter our predicted signals with a 2\textsuperscript{nd} order Butterworth band pass filter with cutoff frequencies at 0.75 and 2.5\,Hz (corresponding to 45 to 150 beats-per-minute) as used in previous work~\cite{liu2020multi}.
Afterward, we calculate the heart rate for each sliding window by taking the frequency with the highest spectral power obtained from the Fast Fourier transform.

\section{Results}
\label{sec:results}

\subsection{Quantitative Analysis}
\label{sec:quantitative_analysis}

\begin{figure}
  \centering
    \includegraphics[width=0.48\textwidth]{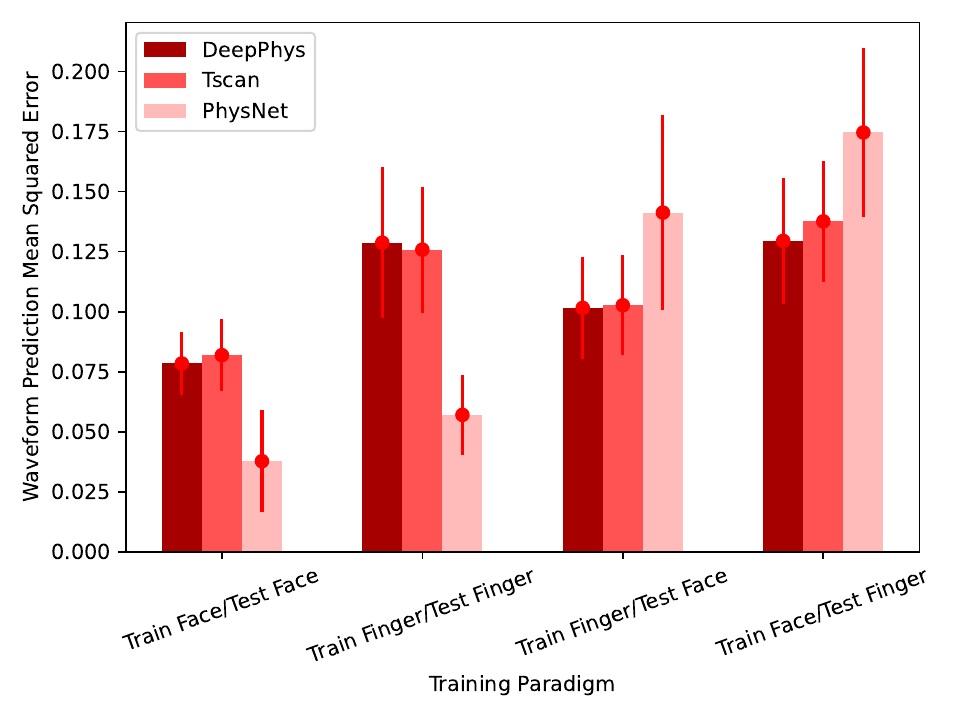}
  \caption{\textbf{Test Loss on Waveform Predictions.} Mean squared error (MSE) between the predicted PPG waveforms and reference contact sensor measurements. Error bars show the standard deviation across subjects. MSE Lower = Better}
  \label{fig:MSE}
\end{figure}

\autoref{tab:mse_results} (and \autoref{fig:MSE}) show the mean squared error (MSE) between the labels and rPPG signals for the four scenarios of training and testing with the face and finger PPG signals as targets.
The lowest MSE for all three models across all participants was obtained when training and testing using the PPG signal from the forehead with MSEs between 0.038 and 0.082 compared to MSEs of 0.057 to 0.129 when training and testing using the finger PPG sensor.
When training with the finger PPG signals and testing with the forehead PPG signals, we obtain MSEs, which are in between the MSEs of only using the forehead/ finger PPG signals.
The worst results are achieved when training with the PPG signal from the forehead and testing on the finger PPG signals with an MSE of up to 0.175 across all participants.

\begin{table*}[htb]
    \centering
    \begin{tabular}{lcccc}
    \toprule
         \textbf{Network} & \textbf{Train Face/ Test Face} & \textbf{Train Finger/ Test Finger} & \textbf{Train Finger/ Test Face} & \textbf{Train Face/ Test Finger} \\
         \midrule 
         DeepPhys~\cite{chen2018deepphys} &  \textbf{0.078} & 0.129 &  0.102 & 0.129 \\
         TS-CAN~\cite{liu2020multi} & \textbf{0.082} & 0.126 & 0.103 & 0.138 \\
         PhysNet~\cite{yu2019remote} & \textbf{0.038} & 0.057 & 0.141 & 0.175 \\
         \bottomrule
    \end{tabular}
    \caption{\textbf{Test Loss on Waveform Predictions.} Mean squared error (MSE) between the predicted PPG waveforms and reference contact sensor measurements. Lower = Better}
    \label{tab:mse_results}
\end{table*}

In~\autoref{tab:mae_results}, we report the associated mean absolute error (MAE) in heart rate estimation from the rPPG waveforms when training and testing using the PPG signals from the forehead and when training with the forehead PPG and testing with the finger PPG.
We achieve MAEs between 2.36 (PhysNet) and 3.04 (DeepPhys) beats per minute when training and testing on the forehead.
When training on the forehead PPG and testing on the finger PPG, we achieve even lower MAEs between 1.26 (PhysNet) and 2.06 (DeepPhys).
Both results are comparable to benchmarks on other datasets for the used neural models~\cite{liu2024rppg}.

\begin{table*}[htb]
    \centering
    \begin{tabular}{lcc}
    \toprule
         \textbf{Network} & \textbf{Train Face/ Test Face} & \textbf{Train Face/ Test Finger}  \\
         \midrule 
         DeepPhys~\cite{chen2018deepphys} &  3.04 &  2.06 \\
         TS-CAN~\cite{liu2020multi} & 2.79 & 1.74 \\
         PhysNet~\cite{yu2019remote} & \textbf{2.36} & \textbf{1.26} \\
         \bottomrule
    \end{tabular}
    \caption{\textbf{Test Loss for Heart Rate Prediction.} Mean absolute error (MAE) between the calculated heart rates from the predicted PPG signals and reference contact sensor measurements. MAE Lower = Better}
    \label{tab:mae_results}
\end{table*}

\subsection{Qualitative Analysis}
\label{sec:qualitative_analysis}

\begin{figure*}[htb]
  \centering
    \includegraphics[width=1\textwidth]{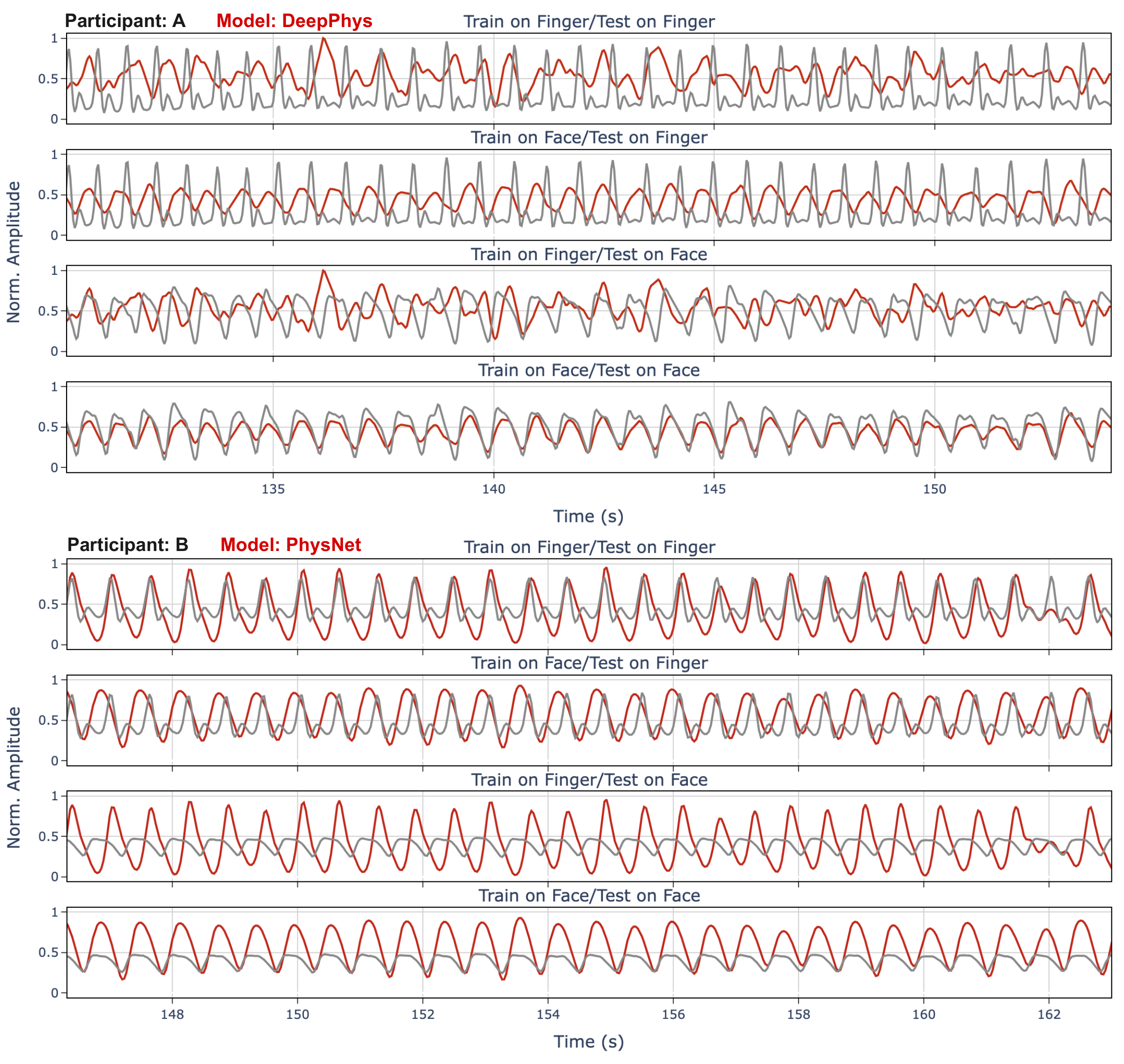}
  \caption{\textbf{Waveform Examples from Two Participants.} Contact sensor (gray) and rPPG predictions (red) for test data from two participants using the DeepPhys model (top) trained on finger and tested on face, (second top) trained on face and tested on finger, (third top) trained on finger and tested on face, (fourth top) trained on face and tested on face.}
  \label{fig:waveforms}
\end{figure*}

Figure~\ref{fig:waveforms} shows qualitative examples of the predicted and ground truth PPG signals under the four different training scenarios.
When training exclusively on facial PPG signals, our neural models achieve the best results, accurately capturing both the temporal dynamics and morphology of the ground truth contact PPG signals. This is best reflected when testing on facial PPG.
However, when using only finger PPG, the models struggle to accurately predict the temporal dynamics and morphology of the reference signal.
While the underlying frequency of the heart rate seems to be represented correctly, the models are not capable of learning the morphology of the finger PPG signal.
Especially noteworthy is also the temporal shift between the predicted and the contact reference signal when training on the face and testing on the finger.
The predicted signal trained on the face video is consistently slightly earlier than the reference PPG signal from the finger.
Generally, we can clearly see how the neural models adapt the morphology of the predicted signal to the waveform of the training reference signal.

\section{Discussion}
\label{sec:discussion}

\subsection{Influence of PPG Site on Training Performance}
\label{sec:performance_analysis}

In our analysis, we found quantitatively and qualitatively that the performance of neural models can be improved when training with PPG signals from the forehead compared to using PPG signals from the fingertip.
The mean squared error between the waveforms of the predicted and reference contact PPG signals decreases up to 40\% when using the PPG signals from the forehead compared to using the PPG signals from the fingertip.
As we see in the qualitative analysis, this has two main reasons, which are both likely caused by the larger domain gap between the input videos of the face and the finger PPG signal compared to the forehead PPG signal.

First, the neural models learn to better predict the morphological characteristics of the PPG waveform derived from the PPG sensor on the forehead than the PPG sensor on the finger.
The neural model only has videos of the face as input and can, therefore, directly learn the morphological characteristics of the forehead PPG waveform.
In contrast, for the finger PPG signal, the neural model additionally needs to learn to map the predicted PPG signal waveform from the forehead to the waveform of the PPG signal from the finger.
This introduces an additional layer of complexity to the learning process. 

As related work has shown, the pulse waves arrive at different times at different locations on the body~\cite{jeong2016introducing, niu2023full}.
At the forehead, the pulse wave typically arrives earlier than on the finger.
Therefore, the neural models also need to learn to map the temporal difference between the pulse arrival in the videos of the face to the pulse arrival compared to the finger PPG sensors.
In Figure~\ref{fig:waveforms} there is a significant difference between the outputs of DeepPhys and PhysNet. We suspect that learning phase shifts in the temporal signal is particularly challenging for a model like DeepPhys that does not model a time component.  Whereas for PhysNet the model is able to learn better waveforms; however, the morphology of the generated waveforms are considerably different from the reference contact measurements.


These results highlight the advantages of using the PPG signal from the same location of the body where the input is obtained from and that the impact of this choice might be different for differnt types of neural models.
By minimizing the domain gap between the inputs and the labels, we show that the performance of state-of-the-art neural models is improved to accurately estimate rPPG signals from videos of the face.
These results can be especially relevant for tasks downstream of the rPPG signals, such as pulse wave analysis based prediction of blood pressure~\cite{jeong2016introducing, bousefsaf2022estimation}.
While remotely predicting blood pressure is still unsolved, the main two approaches at the moment are to either calculate the pulse transit time (PTT) between two different body locations from the rPPG signals or to exploit the morphological characteristics of the predicted rPPG signal.
For both approaches, it is essential that the predicted rPPG signals accurately represent the morphological characteristics of the reference PPG signal and have a time shift/phase offset that is as small as possible.

\subsection{Evaluating rPPG Predictions}
\label{sec:limitations}

When looking at the performance evaluation of related work on rPPG estimation, we notice that most research (for example~\cite{chen2018deepphys, yu2019remote, liu2020multi, yu2022physformer, lee2023lstc}) report  heart rate metrics (such as the MAE or RMSE). We believe that this could be masking the reality that the models do not produce very faithful wavefrom morphology.

Furthermore, we think that our results point toward why unsupervised methods have performed so well. Our results show that the ground truth PPG signals from the finger is not optimal for training supervised methods. 
To improve the performance and to allow for a better comparison of the performance of supervised models, we make two suggestions.
Using PPG signals from the face in the future and also evaluating the model performances using the MSE between the predicted signal and the reference contact PPG signal.
In this way, we can prevent models from only predicting signals with the correct frequency component of the heart rate but the wrong morphology.
This will be especially relevant for future downstream tasks, such as predicting blood pressure, in which morphological characteristics of the predicted rPPG signals are of great importance.

\subsection{Limitations}
\label{sec:limitations}

We see two main limitations of our evaluation in the used dataset.
First, when testing our trained neural models on the finger PPG, we see that the MSE is considerably higher and the qualitative comparison worse compared to testing on the face PPG.
However, we cannot determine with certainty if this is only caused by the domain gap between our input signal (video of the face) and the reference signal (PPG signal from the fingertip).
It is also possible that this is a result of the inherently different waveform of the finger PPG signal (more high-frequency variations), which 
might be more difficult to learn for the neural models.

Second, the dataset size is rather small with $N=18$ participants, has a rather low diversity in skin types, and only includes people between the ages of 19 and 36.
Therefore, we cannot reliably analyze how our results generalize to different skin types or people of different ages.
As the skin thickness decreases when aging, this could impact the generalizability of our results~\cite{branchet1990skin}.
Furthermore, the dataset only has reference contact PPG recordings from the forehead and the fingertip.
With the popularity of wearable devices like smartwatches in recent years, it would have also been interesting to analyze how the PPG waveforms from the wrist differ from the fingertip and face, and how they influence the performance of the trained neural models.
This is especially interesting as some publically available datasets, such as the UBFC-Phys dataset~\cite{meziatisabour2021ubfc}, use a smartwatch (Empatica E4) to record the reference contact PPG signals from the wrist.
Additionally, the used dataset only has one task (self-pinching) and minimizes the motion of the participants' faces by placing the participants' heads on a chin rest.
Therefore, we can also not evaluate how the results generalize to different tasks that, for example, include more motion.
\section{Conclusion}
\label{sec:conclusion}

In this study, we investigated how the performance of different state-of-the-art neural models for rPPG estimation is influenced when training with PPG signals obtained from two different body sites, the forehead, and the commonly used fingertip. 
Our analysis reveals that a significant improvement is achieved when utilizing the PPG signal from the forehead as the ground truth during training compared to using the PPG signal from the fingertip.
We decrease the MSE between the waveforms of the predicted and the ground truth PPG signals by up to 40\% when using the PPG signal from the forehead compared to the fingertip.
We also show qualitatively that the neural models learn the morphological characteristics from the forehead PPG signals better than from the finger PPG signals as we decrease the domain gap between the input and labels.

These results show the importance of considering the placement of the reference contact PPG sensor when designing user studies for rPPG prediction.
Currently, the vast majority of publicly available datasets use contact PPG signals obtained from the fingertip or the wrist.
We hope that in the future, researchers could achieve more accurate physiological signal estimation by leveraging PPG signals from the same location as the input videos, paving the way for improved health monitoring.

{
    \small
    \bibliographystyle{ieeenat_fullname}
    \bibliography{main}
}


\end{document}